\documentclass[twocolumn,aps,prb,preprintnumbers,amsmath,amssymb]{revtex4}
\usepackage{subfiles}
\usepackage{natbib}
\usepackage{ifthen}
\usepackage{color}
\usepackage{graphicx}
\usepackage{epstopdf}
\usepackage{graphicx}
\usepackage{epsfig}
\usepackage{ulem}
\usepackage{latexsym}
\usepackage{amsmath}
\usepackage{amssymb}
\usepackage{bm}
\usepackage{wasysym}
\usepackage{gensymb}
\usepackage{overpic}
\usepackage{latexsym}
\usepackage{amsmath}
\usepackage{amssymb}
\usepackage{bm}
\usepackage{wasysym}
\begin{document}

\title{Emergence of chaos and controlled photon transfer in a cavity-QED network }

\author{Amit  Dey\footnote{amit.dey@icts.res.in}}
\affiliation{International Centre for Theoretical Sciences, Tata Institute of Fundamental Research, Bengaluru -- 560089, India}

\author{Manas Kulkarni\footnote{manas.kulkarni@icts.res.in}}
\affiliation{International Centre for Theoretical Sciences, Tata Institute of Fundamental Research, Bengaluru -- 560089, India}

\date{\today}
\begin{abstract}


We develop optimal protocols for efficient photon transfer in a cavity-QED network. 
This is executed through stimulated Raman adiabatic passage (STIRAP) scheme, where time-varying capacitive couplings (with carefully chosen sweep rate) play a key role. We work in a regime where semiclassical limit is valid and we investigate the dynamical chaos caused by the light-matter coupling. 
We show that this plays a crucial role in estimating the lower bound on the sweep rate for ensuring efficient photon transfer. 
We present Hermitian as well as an open quantum system extension of the model. 
Without loss of generality, we study the three cavity and four cavity case and our results can be adapted to larger networks. Our analysis is also significant in designing transport protocols aimed for nonlinear open quantum systems in general.       
\end{abstract}
\maketitle
\textit{Introduction:}
High-precision controllability of cavity-QED (c-QED) systems and the potential of fabricating artificial lattices \cite{kollar,fitzpatrick,houck,schmidt} highlights c-QED systems as an important component of quantum network \cite{cirac,long,vogell,kato,meher,biswas}. The accessibility of a wide range of light-matter interaction (nonlinearity) signify its relevance for simulating strongly correlated systems \cite{hartmann,mendoza,coto,jin} and demonstrate various phases such as localization-delocalization \cite{ad,tureci, raftery}, superfluid-mott insulator \cite{hartmann,coto,greentree,rossini,aichhorn,brown} phases. Interesting and important phenomena such as qubit state preparation, photon assisted transfer \cite{cirac,vogell,kimble,reiserer} and various quantum correlation measures \cite{reiserer,orszag,kulkarni}, to name a few, have also been recently investigated.  

Population transport through a nonlinear network [such as multi-mode Bose-Hubbard (BH) systems] results in intricate physics of various types of instabilities \cite{ad1,korsch}. Apart from energetic instability \cite{korsch} due to nonlinear eigenstates (of the problem in the semi-classical limit), chaos 
can play major role in determining transfer efficiency \cite{ad1}. 
Therefore, a judicious control of system parameters is crucial to tackle such sensitive physical processes. Nonlinear STIRAP consisting of interacting atomic Bose-Einstein (BEC) condensates has been analyzed semiclassically \cite{ad1} as well as in a quantum many-body framework \cite{ad2}, and the role of various instabilities have been investigated theoretically. It has been shown that the adiabatic conditions for such processes get modified due to emergence of chaos \cite{ad1}. 
Another platform to investigate nonlinearity is a c-QED lattice. This platform precisely implements the Jaynes-Cummings nonlinearity which is very different from the BH nonlinearity in atomic BEC. In addition to this, a dispersive regime of a  c-QED can mimic the BH nonlinearity (\textit{Kerr} type).
Therefore, a c-QED lattice, being an efficient quantum simulator, demands an extensive analysis of nonlinear transport. Chaotic signature in systems where a single cavity is involved\cite{larson,miguel,prants,prants1,prants2} has been investigated and such systems can be considered to be a good testing bed for quantum-classical correspondence \cite{miguel} of chaos. In a linear trimer of cavities\cite{fazio}, control of non-directed (unlike STIRAP scheme) single photon transfer is proposed by tuning the ratio of inter-cavity tunnelings in ultrastrong light-matter coupling regime. Although nonlinear contribution is studied for adiabatic light passage in terms of excitation power dependence \cite{lahini}, to the best of our knowledge, role of chaos in these optical processes remained elusive so far. The Jaynes-Cumming (JC) interaction-induced nonlinearity is exploited in coupled c-QED systems and delocalized-localized phases have already been realized \cite{ad,raftery,tureci}. Furthermore, driven-dissipative preparation of exotic steady states in extended cavity systems paved the avenue of controlling photon propagation in scaled-up architectures\cite{ad}. Therefore, a deeper understanding of aspects of nonlinear dynamics (such as efficient photon transfer) of these systems will significantly add to the existing control strategies and it is much needed to open up myriad of technological applications \cite{kimble}. Developing such protocols warrants a deep understanding of nonlinear systems and subsequently bringing in important notions (for e..g., chaos) can play a paramount role in engineering the systems to ensure efficient transfer. 

In this paper, we investigate a c-QED based STIRAP and show that dynamical chaos sets the lower bound for the sweep rate (which quantifies how fast one tunes the coupling strength), resulting in efficient photon transfer. Without loss of generality, we study the case of three and four cavities, and by efficient photon transfer, we mean, a nearly $100\%$, transfer of photons from the first cavity to the last cavity with almost no occupation of the intermediate cavities during the time evolution.  Quantifying chaos by Lyapunov exponent (LE) in the semi-classical limit, we make connection with the sweep rate. This sets the lower bound on the sweep rate for the tuning parameters and thereby helps in achieving a strategy to ensure nearly $100\%$ transfer of photons in an interacting/nonlinear system. Such a successful strategy for interacting systems lays a strong foundation for (i) establishing photon mediated communication by minimizing dissipation in a quantum network (because intermediate cavities are essentially empty in the process) and (ii) qubit-state transfer and its readout at the terminal cavity.  
%
%

Starting from the model Hamiltonian, we work out the semiclassical equations of motion as we are in a regime where it is valid. 
We obtain the stationary point (SP) solutions for the Hermitian problem at every stage of sweep (see $J_1,J_2$ sweep in Fig.~\ref{fig_schem}). These can be typically multi-valued but we track a special SP (SSP) branch that leads to a near perfect transfer from the first to the last cavity with negligible content in the intermediate cavities. We present the STIRAP time dynamics for various sweep rates and analyze chaotic effects. For SSP branch, we present LE analysis at different sweep stages. This characterises the chaotic aspects of the system.  The results after including the inevitable presence of dissipation in experiments has also been discussed in supplementary material~\cite{supp}. Although much of our analysis relies on a semi-classical approximation, we have successfully demonstrated the consistency with an exact quantum calculation (see supplementary material~\cite{supp}). We then summarise our findings and discuss the future outlook.
\begin{figure}[htb]
  \centering
   \includegraphics[width=2.8in]{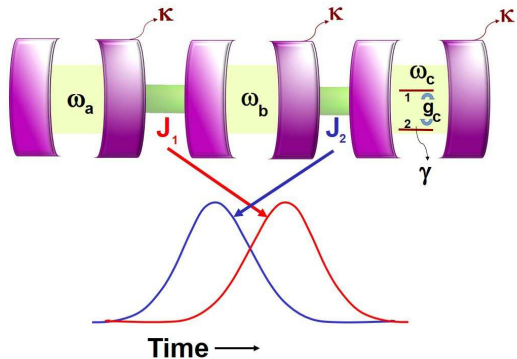}
  \caption{(Color online) Schematic diagram presenting three coupled c-QEDs (with photon frequencies $\omega_{a},\omega_b,\omega_{c}$) acting as a STIRAP. We consider cavity-qubit coupling $g_c$ only for cavity-c, where the levels of the qubit are marked as $1$ and $2$. The objective of transferring photon population from cavity-a to cavity-c by negligibly populating the intermediate cavity-b is achieved through the counter-intuitive sequence of Gaussian pulses for the tunnelling $J_1,J_2$. Without loss of generality, we have shown the three cavity case although our analysis holds for a large network.}
  \label{fig_schem}
\end{figure}\\
\textit{Model and dynamical equations:} 
The c-QED STIRAP given by the schematic Fig.~\ref{fig_schem} can be described by the time-dependent Hamiltonian given by,
\begin{equation}
 \hat{H}=\sum_{j\in \{a,b,c\}}H_j - J_1(t) (\hat{a}^{\dagger}\hat{b}+h.c.) - J_2(t) (\hat{b}^{\dagger}\hat{c}+h.c.)
 \label{ham}
\end{equation}
where $\hat{H}_a=\omega_a \hat{a}^{\dagger}\hat{a}+\Omega_a \hat{s}^z_a+g_a(\hat{a}^{\dagger}\hat{s}^{-}_a+h.c.)$ describes the Jaynes-Cummings Hamiltonian for the cavity labeled by `a' (cavity-a). $\hat{a}$ destroys a photon with frequency $\omega_a$ in cavity-a, $g_a$ denotes light-matter (photon-qubit) interaction in cavity-a. We define $\hat{n}_{a,b,c}$ as the photon number operator in cavity-a, cavity-b, cavity-c respectively. The two-level system with transition frequency $\Omega_{a,b,c}$ is described by the spin operators $\hat{s}^{\alpha}_a$ (where $\alpha \equiv \{x,y,z\}$). 
The time-dependent couplings are defined as Gaussian pulses $J_{1,2} (t)=Ke^{-\big(\frac{t-t_{1,2}}{\tau}\big)^2}$ (with the sequence $ t_1>t_2$), where $\tau$ is the pulse width and therefore $1/\tau$ is the sweep rate (measured henceforth in units of $K$). Therefore, the Hamiltonian (Eq.~\ref{ham}) is explicitly time-dependent in a rescaled time $\tilde{t} \equiv t/\tau$.
Throughout the paper we use $t_1/\tau=3.697, t_2/\tau=2.4242$ and $K=1$. The evolution under the time-dependent Hamiltonian (Eq.~\ref{ham}) depends on the sweep rate of $J_{1,2}(t)$. 
The STIRAP scheme can be implemented, for e.g., in coupled optical waveguides 
through variation of the spacings between the central and terminal waveguides\cite{kenis,torosov,della,longhi,papalakis,lahini}.
Similar variations of couplings in time domain, is also, in principle implementable \cite{roadmap,wallraff}.
Although there is no nonlinear photon-photon interaction in the Hamiltonian, perturbative treatment in dispersive regime shows that the light-matter interaction induces such interaction \cite{blais}. However, we deal with a resonant situation ($\omega_{a,b,c}=\Omega_{a,b,c}$) where the effect of anharmonicity is most manifest. 

For a linear system (with $g_{\text{a,b,c}}=0$), Eq. (\ref{ham}) is just a standard STIRAP Hamiltonian, which ensures the existence of eigenstate $|\Psi_0\rangle \equiv {\rm cos} \Theta |A\rangle -{\rm sin}\Theta |C\rangle $ where ${\rm cos}\Theta=\frac{J_{2}}{\sqrt{J^2_{1}+J^2_{2}}}$. Here, $|A\rangle$ ($|C\rangle$) is the state vector when the total population resides at cavity-a (cavity-c). This particular eigenstate does not project on to $|B\rangle$ and acts as a `dark state' \cite{vitanov}, i.e., $\langle B |  \Psi_0 \rangle = 0$. However, as $\Theta$ varies from 0 to $\pi/2$ (i.e., a complete sweep of $J_{1,2}$, see Fig.~\ref{fig_schem}), the state vector $|\Psi_0\rangle$ rotates from $|A\rangle$ to $|C\rangle$ resulting population transfer.  We define transfer efficiency as $T=\frac{\langle \hat{n}_c(t\to\infty)\rangle}{\langle \hat{n}_a(t=0)\rangle}$. A semiclassical analog of an eigenstate is a solution of semi-classical equations of motion that do not evolve (i.e., a SP solution). Since, we will work in a regime where semi-classical approximation is valid, the analog of the dark state mentioned above is a SSP branch that leads to a perfect transfer from cavity-a to cavity-c (with zero content in cavity-b throughout the time dynamics). Therefore, for a successful adiabatic passage, we sweep the couplings  $J_1, J_2$ slow enough (i.e, $1/\tau$ is relatively small), so that the time evolution of the quantities follows the SSP branch.

In this paper, we deal with a more complicated nonlinear situation and seek similar photon transfer mechanism. 
In the large photon number limit, we exploit the approximation $\langle \hat{a} \hat{s}^+_a \rangle\approx \langle \hat{a}\rangle \langle \hat{s}^+_a \rangle$ and treat the problem in a semiclassical framework~\cite{tureci,kulkarni_dqd}. For brevity, we use the following notations: $\langle \hat{a} \rangle \rightarrow a$, $\langle \hat{s}^-_a\rangle \rightarrow s_a$, and $\langle \hat{s}^z_a \rangle \rightarrow s^z_a$. Using Heisenberg equation of motion and incorporating the above approximation we obtain the dynamical equations given by the following five equations, 
\begin{eqnarray}
 \dot{a}&=&i J_{1} b, \label{eom1}\,\,
 \dot{b}=-i\Delta b +i[J_{1} a + J_{2} c],\label{eom2} \\
 \dot{c}&=& -ig_c s_c+iJ_{2} b,\label{eom3}\\
 \dot{s}_c&=& 2ig_c c s^z_c,\label{eom4}\,\,
 \dot{s}^z_c= -ig_c[{s^*_c} c - c^* s_c].
 \label{eom5}
\end{eqnarray}
In obtaining Eqs. (\ref{eom1})-(\ref{eom5}) we consider a setup where only third cavity has qubit-cavity coupling, i.e., $g_a=g_b=0, g_c\neq 0$. This is because, we want $\{n_a,n_b,n_c,s_c,s_c^{z}\} \equiv \{N,0,0, 0,-0.5\} $ to be a SP at $\tilde{t} =0$ which is experimentally more feasible. The more general case, $g_{\text{a,b,c}}\neq 0$ will have more complicated SP solutions which at  $\tilde{t} =0$ may not have easily experimentally implementable initial conditions. 
%
In addition, we consider $\omega_a=\omega_c=\omega_b-\Delta$ without loss of generality and write Eqs. (\ref{eom1})-(\ref{eom5}) in the rotating frame of frequency $\omega_{a,c}$. Here, $\Delta$ is the detuning of cavity-b from cavity-a/cavity-c. 
We work with the Hermitian problem governed by Hamiltonian (Eq.~\ref{ham}), subjected to the conservation $n_a+n_b+n_c+s^z_c+1/2=N$. We show results pertaining to non-Hermitian processes in the supplementary material \cite{supp}. It turns out that, the non-Hermitian results for a complete sweep reflect similar features as the Hermitian case, provided the dissipation rates are considerably less than $1/\tau$.\\

\textit{Stationary Point Solutions:}
\begin{figure}[b]
  \centering
   \includegraphics[width=3.5in]{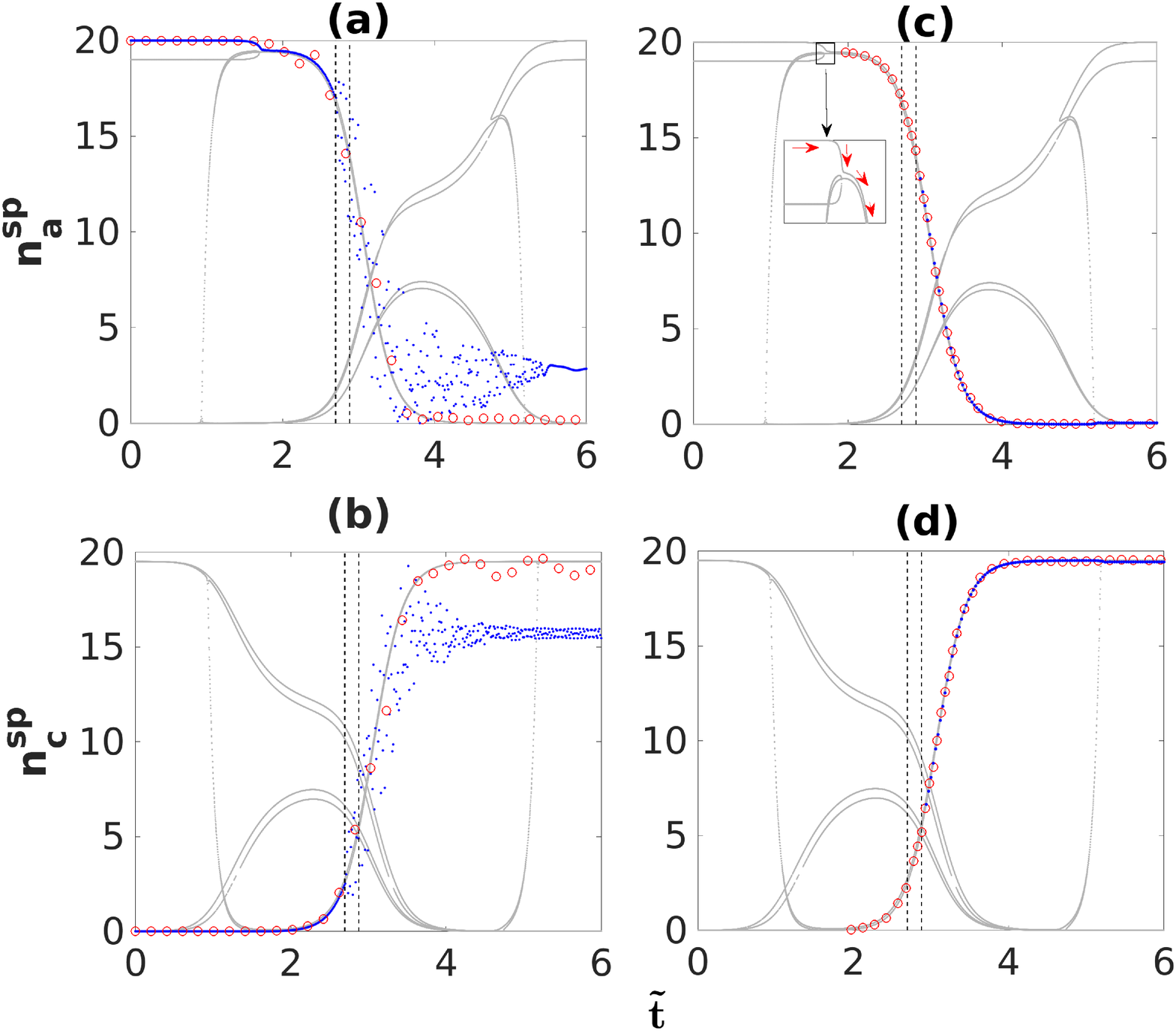}
  \caption{(Color online) The SP solutions for $g_c=0.2K$ are presented by grey lines in all four panels. Red circles and blue dots give the real-time dynamics of cavity populations (under the time dependent Hamiltonian in Eq.~\ref{ham}) at sweep rates $1/\tau=0.0202$ (faster) and $1.2121\times 10^{-4}$ (slower), respectively. The left and right vertical lines (at a certain $\tilde{t}_{\text{left}}$ and $\tilde{t}_{\text{right}}$) are predicted by the LE analysis of Fig.~\ref{fig3} and Fig.~\ref{fig4} (to be discussed later). These vertical lines give us the window of parameters  ($\tilde{t}_{\text{left}}<\tilde{t}<\tilde{t}_{\text{right}}$)  of the Hamiltonian where the system is chaotic. 
 (a) and (b) are respectively,  cavity-a and cavity-c populations, when the system is initialized at the SP solution at $\tilde{t}=0$ (i.e., $\{n_a,n_b,n_c,s_c,s_c^{z}\} \equiv \{20,0,0, 0,-0.5\} $).  On the other hand, in (c) and (d), we present two cases both in which we initiate the system at $\tilde{t}>0$ on the SSP branch. One case is in which the system is initiated immediately after the location of the inset in Fig.~\ref{fig1}c (i.e., red circle, initiated at $\tilde{t}=1.9697$ with SSP solution $\{n_a,n_b,n_c,s_c,s_c^{z}\} \equiv \{19.4542,0.0150,0.0396, 0.4999,-0.0088\} $) and swept at faster rate. Another case is when the system is initiated immediately after the chaotic window (i.e., blue dots, initiated at $\tilde{t}=2.9394$ with SSP $\{n_a,n_b,n_c,s_c,s_c^{z}\} \equiv \{12.8592,0.0073,6.6399, 0.4999,-0.0065\} $) and swept at slower rate. The inset in  Fig.~\ref{fig1}c shows the regime where SSP branch (marked by red arrows) shows 
 a sharp change in SP solutions.}
  \label{fig1}
\end{figure}
The SP solutions are parametrised by $\tilde{t}$ and are therefore independent of $\tau$.  To obtain SP solutions at various $\tilde{t}$, we define the quantity $\hat{h}\equiv\hat{H}(\tilde{t}) - \mu(\tilde{t}) (\hat{n}_a+\hat{n}_b+\hat{n}_c+\hat{s}^z_c+1/2)$ where $\mu(\tilde{t})$ is a Lagrange multiplier (chemical potential) ensuring conservation. 
As in previous section, we derive Heisenberg equations of motion with respect to $\hat{h}$ and by setting the time-derivatives to zero, we write below four equations,
\begin{eqnarray}
 J_1b+\mu a&=&0\label{sp1}, \,\,
 \Delta b-J_1 a-J_2 c-\mu b=0, \label{sp2}\\
 J_2 b-g_c s_c+\mu c&=&0, \label{sp3}\,\,  2g_c c s^z_c+\mu s_c=0. \label{sp4}
\end{eqnarray}
The above four equations along with the constraint $n_a+n_b+n_c+s^z_c+1/2=N$ and the fact that $\frac{s_c^2 + s_c^{*^2}}{2}+s_c^{z^2} = 1/4$ (spin length) gives us six equations and six unknowns ($a,b,c,s_c,s_c^z,\mu$). This has been solved numerically to yield SP solutions at respective $\tilde{t}$ values (Fig.~\ref{fig1}). In particular, we look for SSP branch that facilitates cavity-a to cavity-c transfer by sweeping $\Theta (\tilde{t})$ from $0$ to $\pi/2$.
\begin{figure}[b]
  \centering
   \includegraphics[width=3.5in]{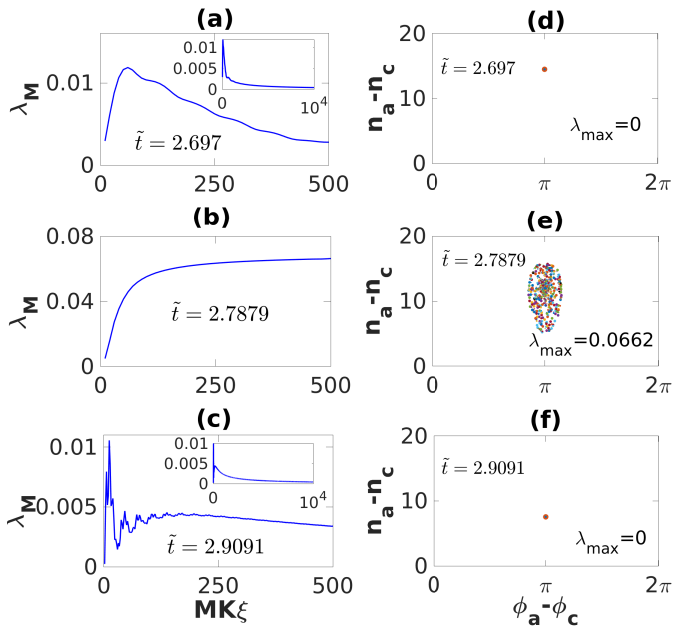}
  \caption{(Color online)(Left column) Lyapunov exponents for $g_c=0.2 K$.  The insets in (a) and (c) demonstrate the vanishing of LE at sufficiently long times as one would expect. (Right coloumn) The dynamics of an ensemble of phase space trajectories. An ensemble of $10$ samples is initiated around the SSP at that particular $\tilde{t}$  and evolved under time-independent Hamiltonian $H(\tilde{t})$ for a long time. As shown clearly, whenever the Lyapunov exponent remains positive [see (c)], the phase space points spread out [see (e)] of the SSP demonstrating the chaos.}
  \label{fig3}
\end{figure}\\

\textit{Numerical Results:}\label{simulation}
In this section, we show the SP solutions for $g_c = 0.2 K$ and present the time dynamics for various sweep rates $1/\tau$ (Fig. \ref{fig1}). We  present SP solutions only for $n^{\rm SP}_a$ and $n^{\rm SP}_c$ since we are interested in photon number of source (cavity-a) and terminal cavity (cavity-c). Among the various SP branches, there is a SSP branch that starts as \{$n^{\rm SP}_a=N=20$, $n^{SP}_c=0$\} at $\tilde{t}=0$ and ends as \{$n^{\rm SP}_a \approx 0$, $n^{\rm SP}_c\approx N$\} with negligible value of $n^{\rm SP}_b$. This special branch is the nonlinear analog of `dark state'.  This SSP branch helps us to choose the correct initial conditions $ \{ a (0),b (0),c (0),s_c (0),s_c^z (0) \}$ that need to be subjected to Eqns. \ref{eom1}-\ref{eom5} for a chosen value of pulse width ($\tau$) in $J_1(t)$ and $J_2(t)$. In addition, we need to make sure that the time dynamics remains on the SSP branch. This can be achieved by choosing optimal (elaborated below) sweep rate $1 / \tau$ subsequently leading to efficient photon transfer. 

Fig.~\ref{fig1} captures two main aspects. One is the SP solutions which are dependent only on $\tilde{t}$. The other aspect is the real time dynamics (which depends on sweep rate $1/\tau$) where one wishes it to adiabatically follow the SSP branch that leads to efficient photon transfer. It is worth recapping the adiabatic theorem which states that upon slowly varying the parameters of the Hamiltonian, the system remains in the Hamiltonian's instantaneous eigenstate. We adapt a similar intuition for its semi-classical limit which forms the basis for the standard STIRAP scheme \cite{vitanov}.  This implies that sweep rates cannot be too fast irrespective of whether the Hamiltonian is interacting or non-interacting. On the other hand, the sweep rates cannot be too slow if the system is interacting (chaotic to be more precise) as demonstrated in Fig.~\ref{fig1} and caption therein. Fig. \ref{fig1} (a) and (b) demonstrate near perfect transfer when the couplings are swept relatively faster (red circles). A relatively slow sweep (blue dots) shows smooth following of SSP branch only till the onset of chaos. Our findings therefore demonstrate that the standard notion of adiabaticity is contradicted when the system is chaotic~\cite{ad1}. To ensure a smooth following of the SSP branch even in the chaotic window, the sweep rate should be sufficiently faster. To show that chaos is the only origin of this breakdown, we show in Fig.~\ref{fig1} (c) and (d) that if the system is initiated by avoiding the chaotic region (shown as blue dots, slower sweep), then it follows the SSP branch. Therefore, the standard notion of adiabaticity holds here.  In the inset region of Fig.~\ref{fig1}(c), one can notice a sharp change in the SP solutions. To follow the SSP branch in this region the sweep rate needs to be slower than the rate of change of of the corresponding energy of the SSP solution w.r.t., $\tilde{t}$. This is the reason why we have small deviations from SSP branch in Fig.~\ref{fig1}(a) for the faster sweep rate case (red circles). The same faster sweep gives a smooth following of SSP branch [Fig.~\ref{fig1}(c) and (d)] if we initiate the system after the inset region. We have also done analysis for higher $g_c$ values and our findings still holds (see supplementary material \cite{supp}).  

\begin{figure}[htb!]
  \centering
   \includegraphics[width=3.3in]{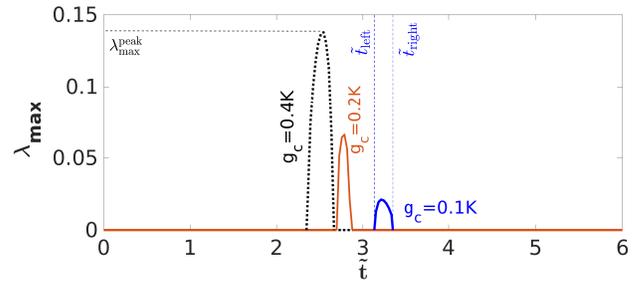}
  \caption{(Color online)  $\lambda_{\rm max}$ is plotted w.r.t. $\tilde{t}$ for $g_c=0.1K$ (solid blue or dark), $g_c=0.2K$ (solid orange or grey), and $g_c=0.4K$ (dotted black). Stronger light-matter coupling implies larger $\lambda_{\rm max}^{\text{peak}}$.}
  \label{fig4}
\end{figure}
\textit{Lyapunov exponent analysis:} 
In this section, we will do a LE analysis that will quantify chaos and this will turn out to be an important ingredient in executing the STIRAP scheme explained earlier. One starts with infinitesimally separated initial conditions and extracts LE
from diverging copies of trajectories. 
Due to the fact that  $n_a+n_b+n_c+s^z_c+1/2=N$, our phase space is bounded. 
Consequently, the distance between the trajectories does not grow monotonically and at long times this  may produce false estimate of LE. To circumvent this issue we exploit the prescription of resetting the phase space distance between trajectories in Refs.~\onlinecite{scotti,Benettin,Benettin2,Benettin1,fine}. The method is described as follows: 

Two phase-space trajectories are chosen, so that they initially differ by phase-space distance $\delta_0$. The trajectories are allowed to diverge for a time step $\xi$ and the new distance between the trajectories $\delta_1$ is reset to $\delta_0$. This procedure is repeated $M$ times where $M$ is large. The LE is then defined as, 
$\lambda_M=\lim_{\delta_0\to 0}\frac{1}{MK\xi}\sum^M_{j=1} {\rm log}\Big(\frac{\delta_j}{\delta_0}\Big) $
where $\lambda_M$ means that we have computed the Lyapunov exponent upto the time $t = M\xi$. Here, $\delta_j$ denotes the deviation between the trajectories before the $j^{th}$ reset.  
The maximum LE ($\lambda_{\rm max}$) can be obtained by taking the limit $M\to \infty$ and a positive (zero) $\lambda_{\rm max}$ indicates a chaotic (non-chaotic) behavior. Resetting at every step ensures that the deviation of trajectories is well within the phase space boundary. 

In Fig. \ref{fig3}, in the left coloumn, we plot $\lambda_M$ as a function of dimensionless time $KM\xi$. Each of the figure  is for a representative $\tilde{t}$ value. 
It is to be noted that specifying a $\tilde{t}$, fixes the parameters of the Hamiltonian $J_1,J_2$ therefore making the Hamiltonian explicitly time independent.  
The procedure we employ to generate  Fig.~\ref{fig3}a - Fig.~\ref{fig3}c is the following: We create two infinitesimally seperated copies, $A, B$ such that $P^{(A)}= P^{\rm SP}(\tilde{t})$ and $P^{(B)}= P^{\rm SP}(\tilde{t})  +\delta P(t)$ where $P$ denotes the set $\{a, b, c, s_c, s^z_c\}$. 
It is worth re-emphasizing that $H(\tilde{t})$ is explicitly time-independent for a particular $\tilde{t}$. 
 As can be seen in  Fig.~\ref{fig3} (left coloumn), we see non-chaotic (Fig. \ref{fig3}a), chaotic (Fig. \ref{fig3}b), and again a non-chaotic (Fig.~\ref{fig3}c) behaviour. This is also reflected in the spreading features of phase-space points (right coloumn of Fig. \ref{fig3}). Keeping in mind, that we are interested in a transfer of photons from cavity-a to cavity-c, as a section of our phase space, we choose $n_a-n_c$ on the y-axis and the conjugate variable $\phi_a-\phi_c$ on the x-axis for the right coloumn of Fig. \ref{fig3}. 
 Fig. \ref{fig3} (e) suggests that such chaotic stages should be crossed quickly (by choosing a sufficiently fast sweep rate) to minimise the effect of chaos. 
For higher $g_c$, the analog of Fig.~\ref{fig3} shows higher Lyapunov exponents and more prominent phase-space spreading (see supplementary material\cite{supp}). 

In Fig.~ \ref{fig4}, we show the maximum LE ($\lambda_\text{max}$) as a function of $\tilde{t}$ for three values of $g_c$. As can be seen, for each $g_c$, there is a window $\tilde{t}_{\text{left}}<\tilde{t}<\tilde{t}_{\text{right}}$ where $\lambda_\text{max}>0$. Therefore, LE analysis plays a paramount role in obtaining a window of $\tilde{t}$ where the system is chaotic. In particular, for the case of $g_c = 0.2K$, the vertical lines in Fig.~\ref{fig1} are obtained by the above analysis.
\begin{figure}[t]
  \centering
   \includegraphics[width=3.5in]{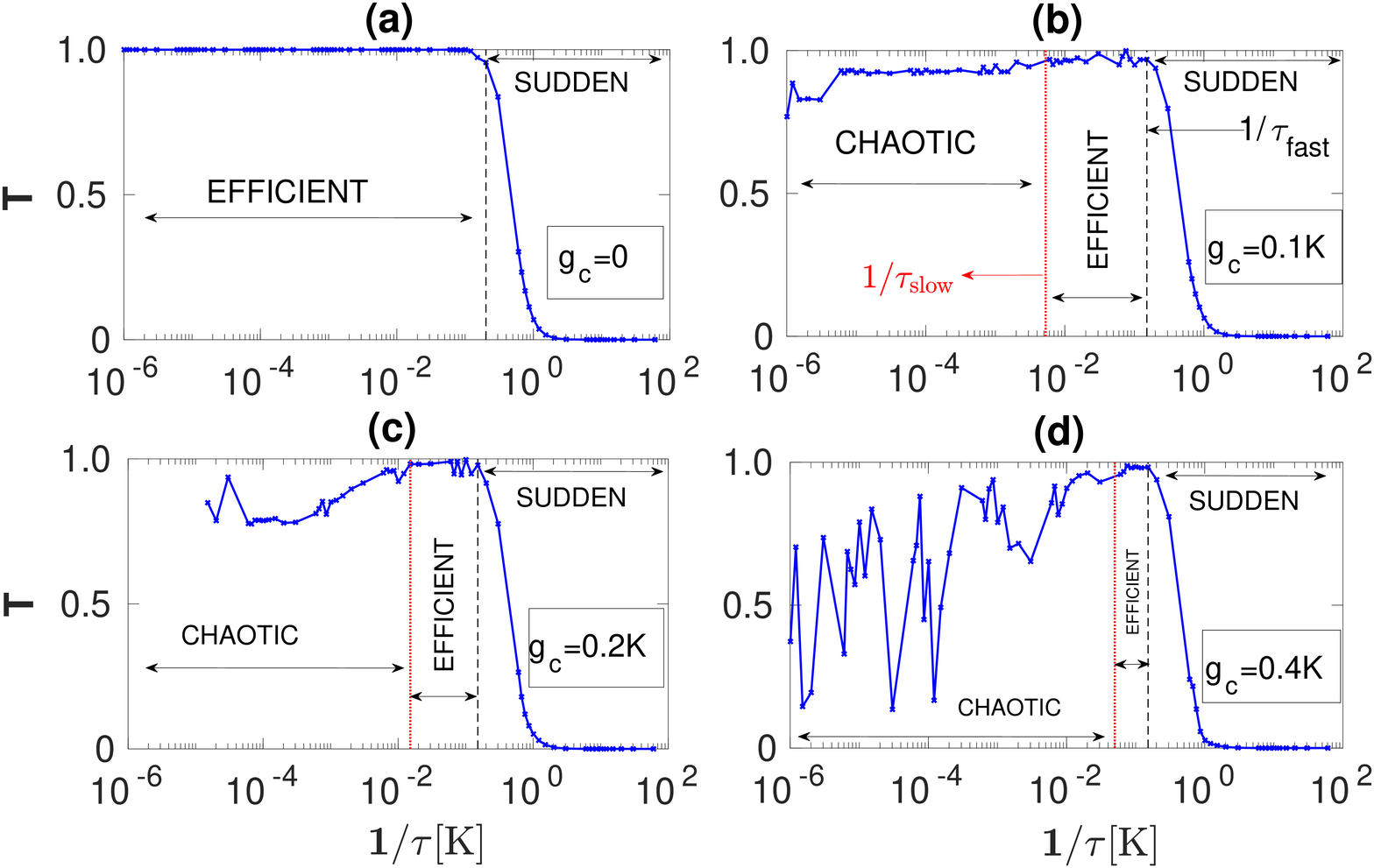}
  \caption{(Color online) T vs $1/\tau$ plot. 
  The black vertical line marks the standard adiabaticity requirement (fastest sweep rate possible) to achieve 95\% efficiency. Red vertical dotted line represents the slowest sweep rate possible so as to assure atleast $95\%$ efficiency. 
  }
  \label{fig5}
\end{figure}

\textit{Transfer efficiency:} 
In Fig.~\ref{fig1} we showed that the dynamics and the transfer efficiency have strong dependence on the sweep rate. In Fig.~\ref{fig5} we plot the transfer efficiency $T$
w.r.t $1/\tau$ for various $g_c$.
Fig.~\ref{fig5} (a) demonstrates that the slow sweep boundary does not exist for noninteracting case, implying no presence of chaos. However, 
the sweep must not be too fast which will result in violation of adiabaticity. 
This feature is reflected in the low-transfer \textit{sudden} region beyond the black dashed lines in Fig.~\ref{fig5}. The black dashed lines ($1/\tau_{\text{fast}}$) in Fig.~\ref{fig5} are constructed such that $T$ becomes less than $0.95$ beyond the line. It is to be noted that $95\%$ is generally regarded as satisfactory high efficiency \cite{vitanov}. In Fig.~\ref{fig5}(b,c,d), we show the interacting case ($g_c\neq 0$). 
Compared to the non-interacting case [Fig. \ref{fig5} (a)] additional chaos-dominated region emerges for the interacting case ($g_c\neq0$). The dotted red vertical line ($1/\tau_{\text{slow}}$) sets a lower bound on the sweep rate below which $T<0.95$. In other words, the sweep rate for high efficiency needs to satisfy $1/\tau_{\text{fast}} >1/\tau > 1/\tau_{\text{slow}}$. It is to be noted that during real time dynamics, the time spent within the chaotic window $(\tilde{t}_{\text{right}}- \tilde{t}_{\text{left}})\tau$ is finite. This means that the relevant LE for chaotic spreading is the finite time LE (see Fig.~\ref{fig3}).  
This automatically implies, $1/\tau_{\text{slow}} <  \lambda^{\text{peak}}_{\text{max}}$ where $ \lambda^{\text{peak}}_{\text{max}}$ is the peak of $\lambda_{\text{max}}$ for a given $g_c$ (see Fig.~\ref{fig4}). 
Therefore, this finding relates LE to the lower bound on sweep rate.
%
%
For higher $g_c$, $1/\tau_{\text{slow}}$ increases thereby shrinking the \textit{efficient} region and widening the \text{chaotic} regime. 
\begin{figure}[t]
  \centering
   \includegraphics[width=3.5in]{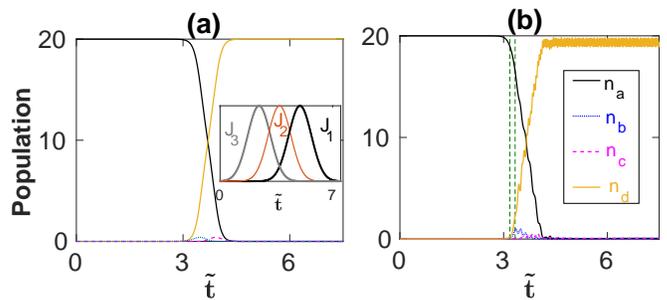}
  \caption{(Color online)  Population dynamics for a four-cavity network with counter-intuitive pulse sequence shown in the inset of (a). Dynamics of $n_a$ (black solid), $n_b$ (dotted blue), $n_c$ (dashed magenta), and $n_d$ [solid yellow (grey)] are shown for (a) linear ($g_{a,b,c,d}=0$) case with $1/\tau=0.00379$ and (b) nonlinear ($g_{a,b,c}=0,g_d=0.2K$) network with $1/\tau=0.0101$. The dashed vertical lines in (b) marks the chaotic window whose potentially detrimental effect (for successful photon transfer) can be overcome with a faster sweep. For both (a) and (b), the system is initialized at the SSP solution of the four cavity network at $\tilde{t}=0$ (i.e., $\{n_a,n_b,n_c,n_d,s_d,s_d^{z}\} \equiv \{20,0,0, 0,0,-0.5\} $).}
  \label{fig7}
\end{figure}

{\it Four-cavity STIRAP scheme:} 
To demonstrate scalability, we extend the three-cavity network to a four-cavity network (cavity-a, cavity-b, cavity-c, cavity-d) where cavity-d houses a qubit. We couple the cavities by counter-intuitive tunneling sequence shown in the inset of Fig.~\ref{fig7} (a). Fig.~\ref{fig7} (a) shows near-unity transfer from cavity-a to cavity-d for a linear closed-system case. Fig.~\ref{fig7} (b) demonstrates near perfect efficiency for the nonlinear case ($g_d \neq 0$) when the parameters are swept at a faster rate. For a slower sweep rate (not shown here), similar to the three cavity case [blue dots in Fig.~\ref{fig1} (a) and Fig.~\ref{fig1} (b)], we find that the efficiency is hindered by chaos. 

%
%
%
%


\textit{Conclusions and Outlook:} 
We have demonstrated a protocol for achieving high transfer efficiency in an interacting c-QED STIRAP network. Such protocols are far from obvious given the fact that we are dealing with a scalable interacting system. To the best of our knowledge, for the first time, we have found and exploited the deep connection between LE and nonlinear STIRAP schemes. 
While, our protocols are developed on a semi-classical platform, we show that the resulting optimal choice of parameters  
successfully achieve our target for the fully quantum case (see supplementary material \cite{supp}).
Our findings are immensely useful for adiabatic light transfer\cite{duncan1,duncan2}, quantum communication and state transfer in cavity-based quantum networks \cite{kato2,chen,pellizzari} and for nonlinear waveguide optics \cite{vitanov}.

Future outlook includes adapting these protocols in different fields where variety of engineered Hamiltonians are achieved (for e.g., Optomechanics \cite{clerk, marquardt,clerk1,tian}). It is interesting to generalize our scheme to higher dimensional systems \cite{houck} and complex geometries \cite{kollar}.  An open fundamental question is connecting Out-of-time-Ordered Correlator (OTOC) \cite{carlos,cameo} and LE in our STIRAP setup especially because STIRAP is a unique platform to access both chaotic and non-chaotic regimes. \\
%

\textit{Acknowledgements}
We thank Lea Santos, Miguel Bastarrachea-Magnani, Duncan O'Dell and Amit Kumar Chatterjee for useful discussions. M. K. gratefully acknowledges the Ramanujan Fellowship SB/S2/RJN-114/2016
from the Science and Engineering Research Board (SERB), Department of
Science and Technology, Government of India. M. K. also acknowledges
support from the Early Career Research Award, ECR/2018/002085  from
the Science and Engineering Research Board (SERB), Department of
Science and Technology, Government of India. MK would like to
acknowledge support from the project 6004-1 of the Indo-French Centre
for the Promotion of Advanced Research (IFCPAR). M. K. acknowledges
support from the Matrics Grant, MTR/2019/001101 from the Science and
Engineering Research Board (SERB), Department of Science and
Technology, Government of India.

\title{ {\bf Supplementary Material for}\\``Emergence of chaos and controlled photon transfer in a cavity-QED network''}
\author{Amit Dey}
\author{Manas Kulkarni}
\address{International Centre for Theoretical Sciences, Tata Institute of Fundamental Research, Bengaluru -- 560089, India}
\maketitle
\begin{widetext}
\section{Semiclassical-Quantum correspondence}
Here, we make a comparison between the semiclassical and exact quantum many-body treatments of our model subject to a STIRAP protocol. The linear case (non-interacting) in Fig. \ref{fig1s} (a) and the nonlinear (interacting) faster-sweep case in Fig. \ref{fig1s} (b) show remarkably good quantum-classical agreement.
\begin{figure}[htb] 
   \centering
   \includegraphics[width=5in]{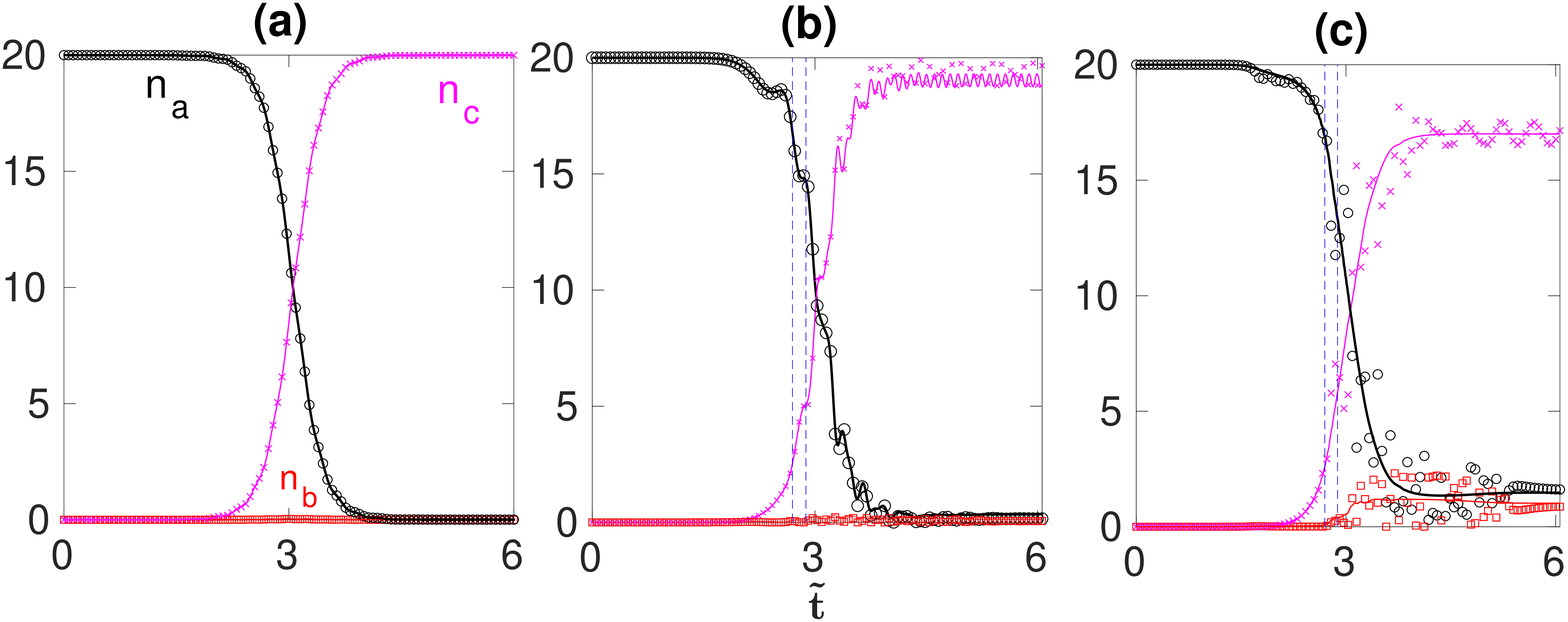} 
   \caption{Population dynamics for (a) $g_{a,b,c}=0$, $1/\tau=0.0303$, (b) $g_{a,b}=0,g_c=0.2K$, $1/\tau=0.0303$, and (c) $g_{a,b}=0,g_c=0.2K$, $1/\tau=0.0012$. Full quantum many-body results are presented by solid lines. The semiclassical plots for $n_a$, $n_b$, and $n_c$ are presented by black circle, red square, and magenta `x' markers, respectively. The vertical dashed lines define the chaotic window as obtained by LE analysis in main text.}
   \label{fig1s}
\end{figure}
In  Fig. \ref{fig1s} (c), both semiclassical and quantum treatments predict decreased efficiency due to chaos for a slower sweep case. In this case [Fig. \ref{fig1s} (c)], we note that the quantum expectation values $\langle n_{a,b,c}\rangle$ do not exactly follow the semiclassical oscillations that begin in the chaotic window (in the slower sweep case). 

In a chaotic region, for quantum and classical counterparts, both the mechanism and diagnostics are different and this is indeed a subject of active interest. This stems from the fact that for the quantum Hamiltonian there are a large number of many body eigenstates (that depends on dimensionality of Hilbert space). However, in a semi-classical approximated description, the number of SP branches are small and this number is solely dependent on the number of solutions of the semi-classical equations of motion. 
Quantum measures of chaos include participation number of eigentates \cite{miguel_sup,ad2_sup}, level spacing statistics \cite{wigner_sup,izrailev_sup}, and 
 Out-of-time-ordered correlator (OTOC) measures \cite{carlos_sup}. The possible analogy between quantum measures, LE analysis of the semi-classical equations in the context of STIRAP scheme is an interesting open question.
 


\section{Dissipative effects on photon transfer}
In main text, we deal with Hermitian dynamics governed by Eq.~1 (of the main text) that neglects various decay channels. In this section, we include photon decay and spontaneous qubit decay quantified by rates $\kappa$ and $\gamma$, respectively. Incorporating dissipation, the modified equations of motion can be written as, 
\begin{eqnarray}
 \dot{a}&=&i J_{1} b-\frac{\kappa}{2}a, \label{eom1d}\,\,
 \dot{b}=-i\Delta b +i[J_{1} a + J_{2} c]-\frac{\kappa}{2}b,\label{eom2d} \\
 \dot{c}&=& -ig_c s_c+iJ_{2} b-\frac{\kappa}{2}c,\label{eom3d}\\
 \dot{s}_c&=& 2ig_c c s^z_c-\frac{\gamma}{2}s_c,\label{eom4d}\,\,
 \dot{s}^z_c= -ig_c[{s_c}^* c - c^* s_c]-\gamma(s^z_c+1/2).
 \label{eom5d}
\end{eqnarray}
The dissipative dynamics leads to a steady state with no photon and the qubit at its ground state. If the dissipation rates are comparable to the sweep rate (i. e., $\kappa,\gamma \sim 1/\tau$) the target goal (of efficient transfer) through STIRAP profile is not achieved. This is because the dissipative effects wash away the possibility of any non-trivial steady state. Therefore, for a practical scenario it is needed that the sweep be completed before the dissipative effects become considerable, i. e., $\gamma, \kappa \ll 1/\tau$.
\begin{figure}[htb]
  \centering
  \includegraphics[width=5in]{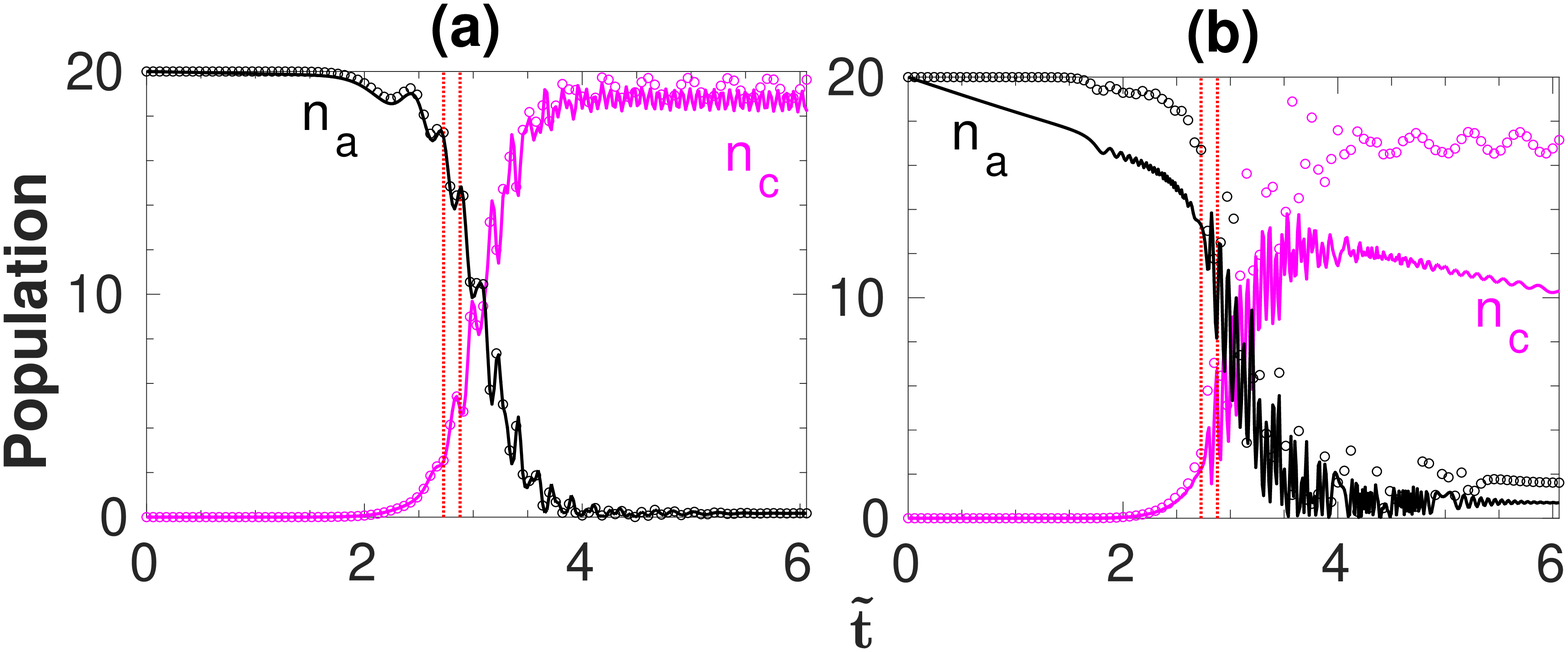}
  \caption{(Color online) Population dynamics (solid lines) for $g_3=0.2 K$ and dissipation parameters $\gamma=\kappa=10^{-4}K$. Dynamics for (a) faster sweep $1/\tau=0.0202$ has no chaotic breakdown but only weak non-adiabaticity introduced at a region where SSP branch changes sharply (as discussed in main text and Fig.~2 therein), (b) slower sweep $1/\tau=0.0012 $ shows breakdown at the $\tilde{t}$ window, where chaos appears for a closed system analysis. The dotted vertical lines are for the Hermitian case and are drawn at same $\tilde{t}$'s as in Fig. 2 (of main text).  
  At steady state all the populations are zero and qubit is in the ground state for both (a) and (b). Here, we show dynamics until one full sweep cycle. For comparision, the Hermitian case ($\gamma=\kappa=0$) is represented by circles for same values of parameters. }
  \label{fig2s}
\end{figure}

\begin{figure}[htb]
  \centering
  \includegraphics[width=3.5in]{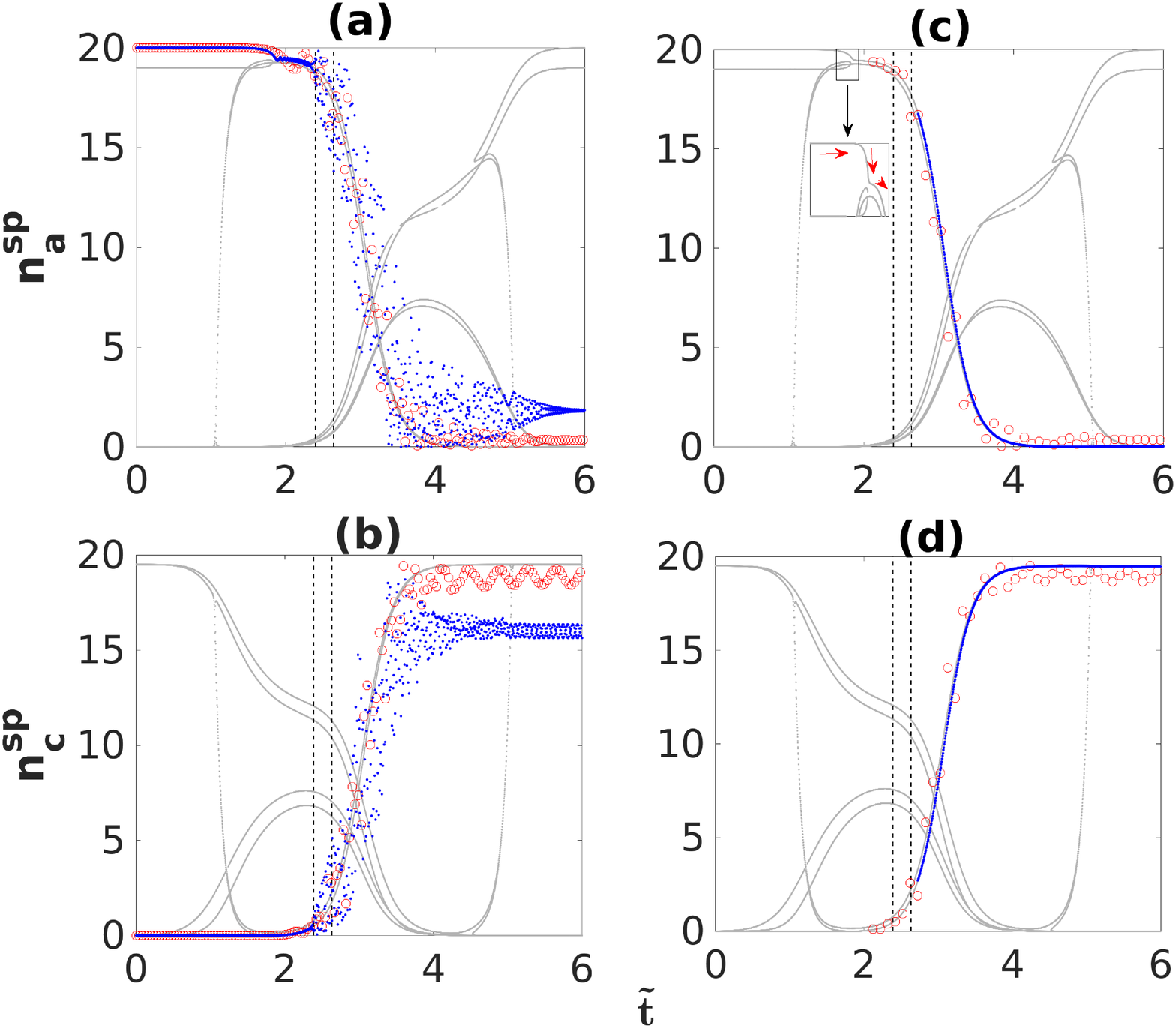}
  \caption{(Color online) The SP solutions for $g_c=0.4K$ are presented by grey lines in all four panels. Red circles and blue dots present the real-time dynamics of cavity populations at sweep rates $1/\tau=.0202$ (faster) and $1.2121\times 10^{-3}$ (slower), respectively. The dashed vertical lines defines the chaotic window. (a) and (b) are respectively,  cavity-a and cavity-c populations, when the system is initiated at the SSP branch at $\tilde{t}=0$. On the other hand, (c) and (d) describe two cases. (i) (red circle) when the system is initiated at SSP branch ($\{n_a,n_b,n_c,s_c,s^z_c\}\equiv \{19.3819,0.0475,0.0798,0.4999,-0.0091 \}$ at $\tilde{t}=2.1212$) after the inset location in (c) and (ii) (blue dots) when the system is initiated at SSP branch ($\{n_a,n_b,n_c,s_c,s^z_c\}\equiv \{16.7613,0.035,2.7105,0.4999,-0.0068 \}$ at $\tilde{t}=2.7273$) just after the chaotic window. Similar to Fig. 2 of main text, the region where the SSP branch changes sharply is zoomed in the inset of (c).}
  \label{fig3s}
\end{figure}

In Fig. \ref{fig2s} (a) we show that a faster sweep offers efficient transport compared to the slower sweep in Fig. \ref{fig2s} (b). Interestingly, the rapid oscillation for slower sweep starts exactly at the place where chaos appears for the Hermitian case in Fig. 1 of main text. As long as the dissipation does not become considerable, the system can still be approximated as a nearly closed system and the Hermitian analysis is valid. Therefore, if $1/\tau \gg \kappa, \gamma$, then for a complete STIRAP cycle, one can make the Hermitian approximation. 




\section{Results for higher $g_c$ value}
Here we present results for $g_{a,b}=0,g_c=0.4K$. 
Compared to the $g_c=0.2K$ case in Fig. 2 (of main manuscript), chaos is more intense (for e.g., as quantified by $\lambda_{\text{max}}^{\text{peak}}$ in Fig.~4 of the main text) and the resulting breakdown appears at comparatively faster sweep rate (than the sweep rate that produces chaotic breakdown in Fig. 2 of main text) in Fig. \ref{fig3s}. This observation is also consistent with Figs. 4 and 5 of main text.
\begin{figure}[t]
  \centering
   \includegraphics[width=4.5in]{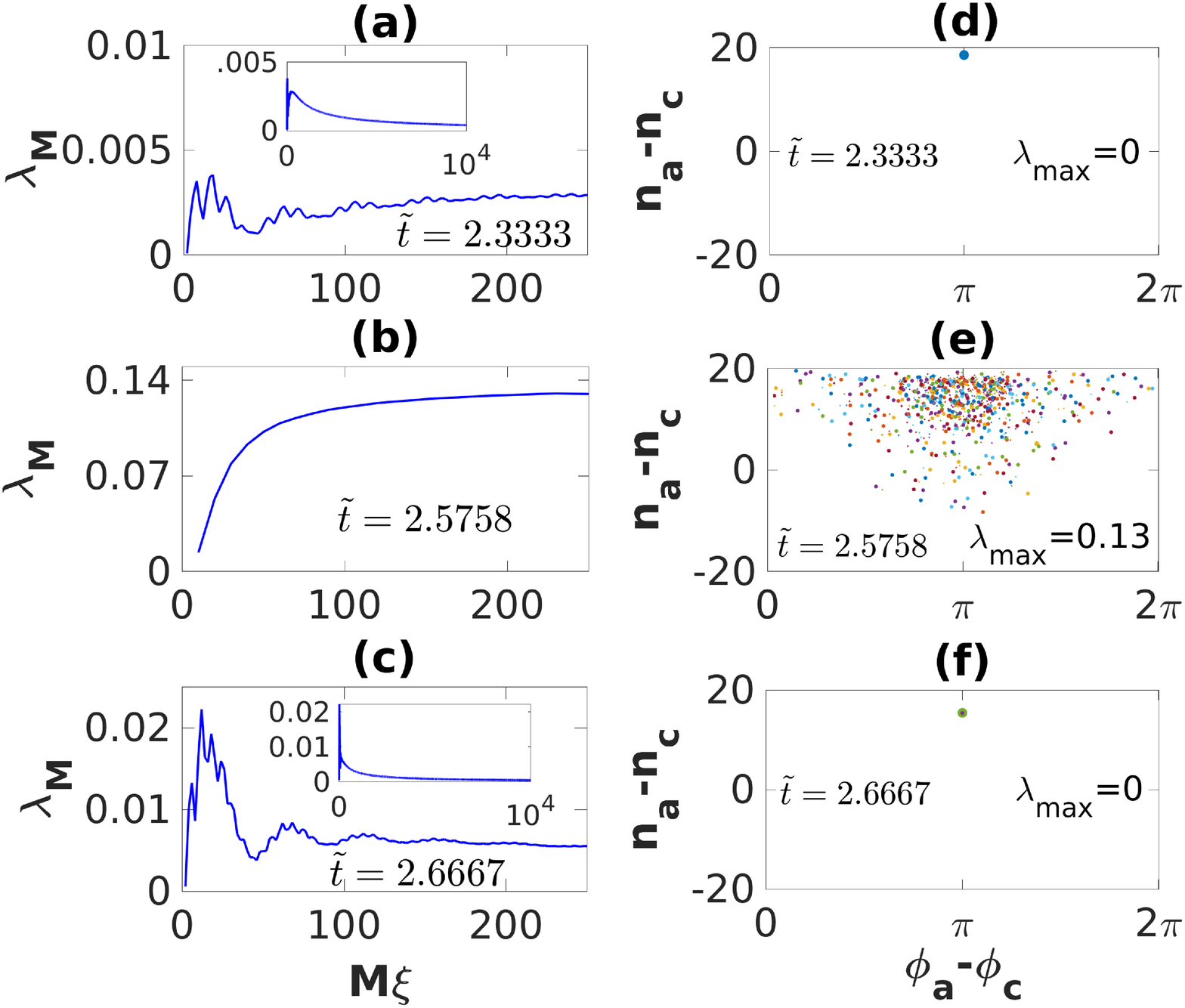}
  \caption{(Color online) Lyapunov exponents are plotted for $g_c=0.4K$.  The insets in (a) and (c) demonstrate the vanishing of LE at sufficiently long times as one would expect.
  The dynamics of an ensemble of phase space trajectories are plotted in (d), (e), (f) for same $\tilde{t}$ values as in (a), (b), (c), respectvely. The ensemble is initiated around the SSP at that particular $\tilde{t}$  and evolved under fixed Hamiltonian $H(\tilde{t})$. Whenever the Lyapunov exponent remains positive [for e.g., in (b)], the phase space points spread out of the SSP confirming the chaotic spreading [for e.g., in (e)].}
  \label{fig4s}
\end{figure}
In Fig. \ref{fig4s} we plot LE for same light-matter couplings as in Fig. \ref{fig3s}. 

\end{widetext}

\end{document}